\def\year{2021}\relax
\documentclass[letterpaper]{article} 
\usepackage{paralist}
\usepackage{aaai212}  
\usepackage{times}  
\usepackage{helvet} 
\usepackage{courier}  
\usepackage[hyphens]{url}  
\usepackage{graphicx} 
\usepackage{natbib}  
\usepackage{caption} 
\usepackage{xspace}
\usepackage{booktabs}
\usepackage{amsmath}
\usepackage{tabularx}

\usepackage[utf8]{inputenc}
\usepackage[T1]{fontenc}
\usepackage{hyphenat}
\usepackage{xspace}
\usepackage{amsmath}
\usepackage{amsfonts}
\usepackage{hyperref}
\usepackage{url}
\usepackage{booktabs}
\usepackage{multirow}
\usepackage{subfigure}
\usepackage{makecell}
\usepackage{caption}
\usepackage{minibox}
\usepackage{bbm}
\usepackage{graphicx}
\usepackage{balance}
\usepackage{mathtools}
\usepackage{color}
\usepackage{marvosym}
\usepackage{ifthen}
\usepackage{textcomp}
\usepackage{enumitem}
\usepackage{verbatim}
\usepackage{algorithm}
\usepackage{algorithmic}
\usepackage{numprint}
\usepackage{balance}

\usepackage{amsthm}
\theoremstyle{plain}

\newcommand{\chatoDisplayMode}[1]{#1}

\definecolor{MyRed}{rgb}{0.6,0.0,0.0} 
\definecolor{MyBlack}{rgb}{0.1,0.1,0.1} 
\newcommand{\inred}[1]{{\color{MyRed}\sf\textbf{\textsc{#1}}}}
\newcommand{\frameit}[2]{
  \begin{center}
  {\color{MyRed}
  \framebox[.9\columnwidth][l]{
    \begin{minipage}{.85\columnwidth}
    \inred{#1}: {\sf\color{MyBlack}#2}
    \end{minipage}
  }\\
  }
  \end{center}
}

\newcommand{\note}[2][]{\chatoDisplayMode{\def\@tmpsig{#1}\frameit{{\Pointinghand} Note}{#2\ifx \@tmpsig \@empty \else \mbox{ --\em #1}\fi}}}
\newcommand{\todo}[2][]{\chatoDisplayMode{\def\@tmpsig{#1}\frameit{{\Writinghand} To-do}{#2\ifx \@tmpsig \@empty \else \mbox{ --\em #1}\fi}}}

\newcommand{\abbrevStyle}[1]{#1}

\newcommand{\ie}{\abbrevStyle{i.e.}\xspace}
\newcommand{\eg}{\abbrevStyle{e.g.}\xspace}
\newcommand{\cf}{\abbrevStyle{cf.}\xspace}

\newcommand{\vs}{\abbrevStyle{vs.}\xspace}

\newcommand{\Secref}[1]{Sec.~\ref{#1}}

\newcommand{\Tabref}[1]{Table~\ref{#1}}
\newcommand{\Figref}[1]{Fig.~\ref{#1}}

\newcommand{\xhdr}[1]{\vspace{1.7mm}\noindent{{\bf #1.}}}

\newcommand{\xhdrNoPeriod}[1]{\vspace{1.7mm}\noindent{{\bf #1}}}

\newcommand{\textcite}[1]{\citeauthor{#1} \shortcite{#1}}

\newcommand{\hide}[1]{}

\makeatletter
\newcommand{\iffont}[2]{\ifthenelse{\equal{\f@family}{#1}}{#2}{}}
\makeatother

\iffont{ptm}{
  \usepackage{mathptmx}

  \DeclareSymbolFont{greek}{OML}{cmm}{m}{n}
  \DeclareMathSymbol{\alpha}{\mathalpha}{greek}{"0B}
  \DeclareMathSymbol{\beta}{\mathalpha}{greek}{"0C}
  \DeclareMathSymbol{\gamma}{\mathalpha}{greek}{"0D}
  \DeclareMathSymbol{\delta}{\mathalpha}{greek}{"0E}
  \DeclareMathSymbol{\epsilon}{\mathalpha}{greek}{"0F}
  \DeclareMathSymbol{\zeta}{\mathalpha}{greek}{"10}
  \DeclareMathSymbol{\eta}{\mathalpha}{greek}{"11}
  \DeclareMathSymbol{\theta}{\mathalpha}{greek}{"12}
  \DeclareMathSymbol{\iota}{\mathalpha}{greek}{"13}
  \DeclareMathSymbol{\kappa}{\mathalpha}{greek}{"14}
  \DeclareMathSymbol{\lambda}{\mathalpha}{greek}{"15}
  \DeclareMathSymbol{\mu}{\mathalpha}{greek}{"16}
  \DeclareMathSymbol{\nu}{\mathalpha}{greek}{"17}
  \DeclareMathSymbol{\xi}{\mathalpha}{greek}{"18}
  \DeclareMathSymbol{\pi}{\mathalpha}{greek}{"19}
  \DeclareMathSymbol{\rho}{\mathalpha}{greek}{"1A}
  \DeclareMathSymbol{\sigma}{\mathalpha}{greek}{"1B}
  \DeclareMathSymbol{\tau}{\mathalpha}{greek}{"1C}
  \DeclareMathSymbol{\upsilon}{\mathalpha}{greek}{"1D}
  \DeclareMathSymbol{\phi}{\mathalpha}{greek}{"1E}
  \DeclareMathSymbol{\chi}{\mathalpha}{greek}{"1F}
  \DeclareMathSymbol{\psi}{\mathalpha}{greek}{"20}
  \DeclareMathSymbol{\omega}{\mathalpha}{greek}{"21}
  \DeclareMathSymbol{\varepsilon}{\mathalpha}{greek}{"22}
  \DeclareMathSymbol{\vartheta}{\mathalpha}{greek}{"23}
  \DeclareMathSymbol{\varpi}{\mathalpha}{greek}{"24}
  \DeclareMathSymbol{\varrho}{\mathalpha}{greek}{"25}
  \DeclareMathSymbol{\varsigma}{\mathalpha}{greek}{"26}
  \DeclareMathSymbol{\varphi}{\mathalpha}{greek}{"27}
  \DeclareSymbolFont{otone}{OT1}{cmr}{m}{n}
  \DeclareMathSymbol{\Gamma}{\mathalpha}{otone}{0}
  \DeclareMathSymbol{\Delta}{\mathalpha}{otone}{1}
  \DeclareMathSymbol{\Theta}{\mathalpha}{otone}{2}
  \DeclareMathSymbol{\Lambda}{\mathalpha}{otone}{3}
  \DeclareMathSymbol{\Xi}{\mathalpha}{otone}{4}
  \DeclareMathSymbol{\Pi}{\mathalpha}{otone}{5}
  \DeclareMathSymbol{\Sigma}{\mathalpha}{otone}{6}
  \DeclareMathSymbol{\Upsilon}{\mathalpha}{otone}{7}
  \DeclareMathSymbol{\Phi}{\mathalpha}{otone}{8}
  \DeclareMathSymbol{\Psi}{\mathalpha}{otone}{9}
  \DeclareMathSymbol{\Omega}{\mathalpha}{otone}{10}
  \DeclareSymbolFont{syms}{OML}{cmm}{m}{it}
  \DeclareMathSymbol{\partial}{\mathord}{syms}{"40}
  \DeclareMathAlphabet{\mathbold}{OML}{cmm}{b}{it}
  \DeclareSymbolFont{largesymbols}{OMX}{cmex}{m}{n}

}
\usepackage{hyphenat}
\usepackage{mathptmx}

\newcommand{\sghandle}{\textit{@slpng\_giants\_pt}\xspace}
\newcommand{\SGB}{SGB\xspace}
\title{Analyzing the "Sleeping Giants" Activism Model in Brazil}

\usepackage{minibox}
\newcommand{\affilSize}{8pt}
\newcommand{\authorbox}[3]{
  \minibox[c]{
    {\fontsize{11pt}{11pt}\selectfont{}#1}\\
    {\fontsize{\affilSize}{\affilSize}\selectfont{}#2}\\
    {\fontsize{\affilSize}{\affilSize}\selectfont{}#3}\\
    \vspace{1mm}
  }
}

\author{
\authorbox{B\'arbara Gomes Ribeiro}{barbaragomes@dcc.ufmg.br}{UFMG}
\authorbox{Manoel Horta Ribeiro}{manoel.hortaribeiro@epfl.ch}{EPFL}
\authorbox{Virg\'ilio Almeida}{virgilio@dcc.ufmg.br}{UFMG, Berkman Klein Center}
\authorbox{Wagner Meira Jr.}{meira@dcc.ufmg.br}{UFMG}
}

\begin{document}

\maketitle

\begin{abstract}
\small
In 2020, amidst the COVID pandemic and a polarized political climate, the Sleeping Giants online activist movement gained traction in Brazil.
Its rationale was simple: to curb the spread of misinformation by harming the advertising revenue of sources that produce this type of content. 
Like its international counterparts, Sleeping Giants Brasil (SGB) campaigned against media outlets using Twitter to ask companies to remove ads from the targeted outlets.
This work presents a thorough quantitative characterization of this activism model, analyzing the three campaigns carried out by SGB between May and September 2020. 
To do so, we use digital traces from both Twitter and Google Trends, toxicity and sentiment classifiers trained for the Portuguese language, and an annotated corpus of SGB's tweets. 
Our key findings were threefold.
First, we found that SGB's requests to companies were largely successful (with 83.85\% of all 192 targeted companies responding positively) and that user pressure was correlated to the speed of companies' responses.
Second, there were no significant changes in the online attention and the user engagement going towards the targeted media outlets in the six months that followed SGB's campaign (as measured by Google Trends and Twitter engagement).
Third, we observed that user interactions with companies changed only transiently, even if the companies did not respond to SGB's request.
Overall, our results paint a nuanced portrait of internet activism.
On the one hand, they suggest that SGB was successful in getting companies to boycott specific media outlets, which may have harmed their advertisement revenue stream. On the other hand, they also suggest that the activist movement did not impact the online attention these media outlets received nor the online image of companies that did not respond positively to their requests.
\end{abstract}

\section{Introduction}
\label{sec:intro}

\begin{figure}[t]
\centering
\includegraphics[width=0.5\linewidth]{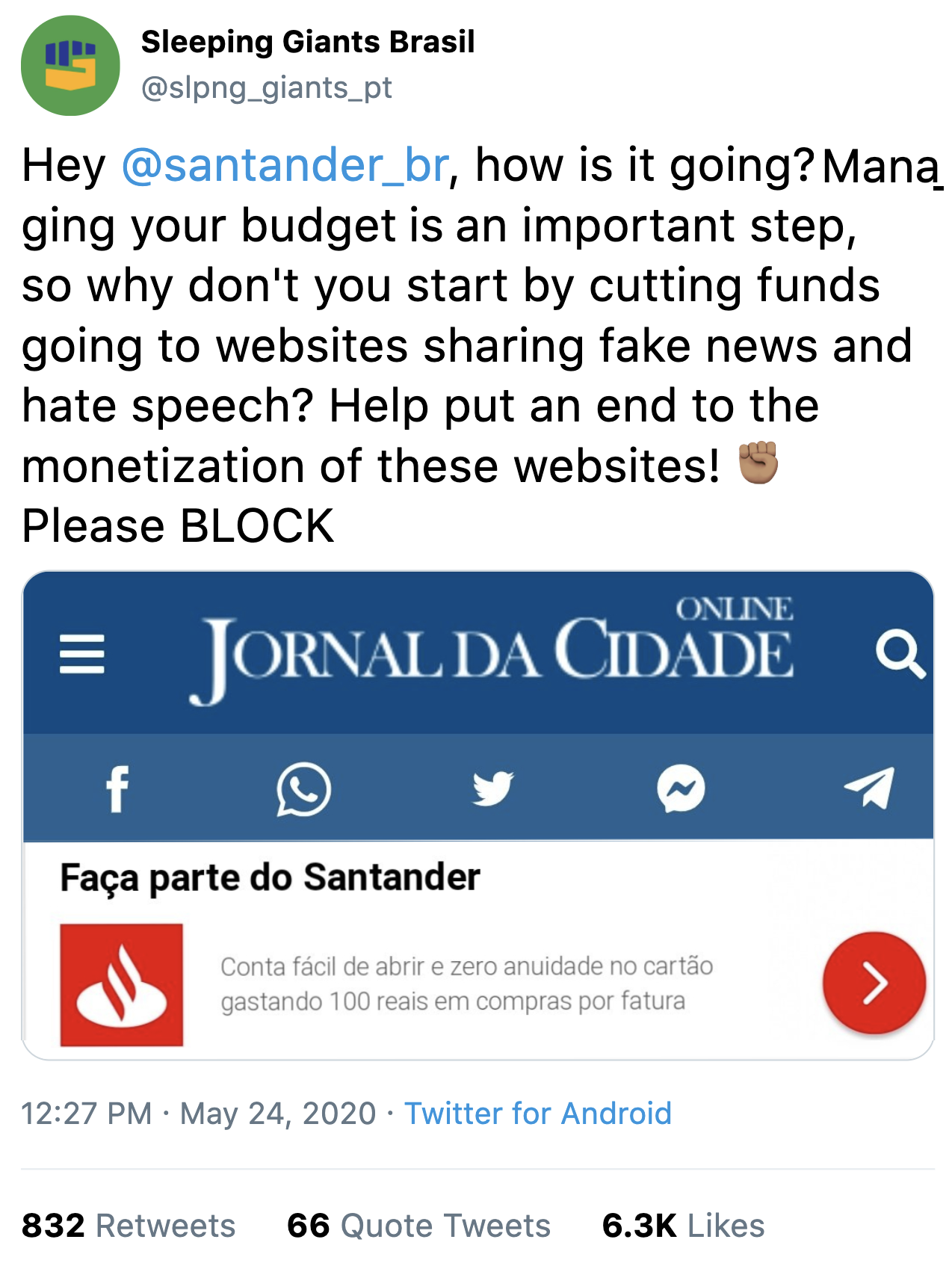}
\caption{Example of Sleeping Giants asking companies to remove ads from a targeted media outlet. Tweet\protect\footnotemark\xspace translated by the authors.}
\label{fig:lead}
\end{figure}

\footnotetext{\url{twitter.com/slpng_giants_pt/status/1264322209922854913}}

Sleeping Giants is an international activist movement centered around Twitter that aims to fight online misinformation using a creative strategy.
Rather than confronting misinformation heads on, for example, fact-checking false claims, the movement tries to persuade companies to remove ads from news outlets that spread hyper-partisan or false information~\cite{braun_activism_2019}. 
Their rationale is clear. 
One of the incentives behind creating and sharing such content is the possibility to monetize it through advertisement~\cite{anderson2017future}.
Thus, harming this revenue stream would be a way to decrease misinformation and improve our information ecosystem.

The initiative started in the U.S. in November 2016 under the Twitter handle \sghandle.
Their first ``campaign'' targeted Breitbart News, a far-right outlet popular among American conservative social media during the 2016 election~\cite{faris_partisanship_2017}.
The movement then spread throughout the globe, with chapters in France (\url{@slpng_giants_fr}), Australia (\url{@slpng_giants_oz}), and most relevant to the work at hand, Brazil (\sghandle).

In Brazil, Sleeping Giants started their activity in May 2020, boycotting outlets spreading misinformation amidst the polarized political climate in the country.
In a month from its creation, their Twitter account (\sghandle) surpassed the original American account, gathering over 400,000 followers~\cite{gq2020}.
Yet, their playbook was the same.
In their first months of existence, they chose three rising media outlets in Brazil known to spread false information consistently (\textit{Jornal da Cidade}, \textit{Brasil Sem Medo}, and \textit{Conexão Política}) and asked over 150 companies (some multiple times) to stop advertising on these websites.
\Figref{fig:lead} illustrates a ``request'' made by \sghandle to a company (\url{@santander_br}), asking them to stop to advertise on \textit{Jornal da Cidade}.

In August 2020, in the context of a lawsuit filed by one of the websites targeted by the movement, a Brazilian judge determined that Twitter should disclose sensitive information related to the account~\cite{sg_zerohora_2020}.
This demand was eventually overturned, but amidst pressure and threats, the creators of the profile revealed their identities to a large Brazilian newspaper~\cite{monica_bergamo_2020}.
These legal hurdles put the Brazilian chapter of Sleeping Giants in the spotlight of Brazilian media for weeks, gathering even more attention towards the activist movement.
With time, \SGB would shift its focus, carrying campaigns to demonetize YouTube channels and influencers, and incentivizing online platforms (\eg, e-commerce and payment services) to curb the spread misinformation and hate speech.

Online activism and its implications have been widely studied in academia (\eg, \citet{christensen_slacktivism, lee_slacktivism}). 
Yet, the Sleeping Giants movement brings something unique to the table: a clearly defined ``activism model'' that has been replicated across the globe and, unlike many, has an objective that is clear and verifiable.

\xhdr{Present work} 
In this paper, we study the "Sleeping Giants" activism model, analyzing how the movement boycotted media outlets in its first months of existence. We use digital traces from 1) Sleeping Giants Brasil (SGB); 
2) the media outlets themselves;
and 3) companies that were advertising in these outlets. 
We ask:

\begin{itemize}[topsep=0pt, labelindent=0pt]
\item \textbf{RQ1: Measuring the success of SGB's requests.} How successful were \SGB's requests for companies to remove their advertisements from targeted outlets? What factors are associated with their success?

\item \textbf{RQ2: Measuring changes to the targeted media outlets.} How did online attention and engagement towards media outlets targeted by \SGB progress in the following months?

\item \textbf{RQ3: Measuring changes to the targeted companies.} After receiving requests by \SGB, did the nature or the volume of online interactions with companies change? Were changes similar for companies that did and those that did not reply to the requests?
\end{itemize}

To answer these questions, we use a carefully annotated corpus of Tweets that captures when \SGB  made requests to companies and how (and if) the companies replied. We complement this data with engagement traces from Twitter (\eg, mentions companies received before and after \SGB's requests) and search volume time-series obtained through Google Trends (to analyze the popularity of the media outlets targeted by \SGB). From the text present in Tweets, we also derive text-based metrics capturing affect\hyp{}related signals (toxicity score, positive/negative sentiment score)~\cite{perspective_api,thelwall_sentistrength_2010}.

\xhdr{Summary of findings} 
Our results suggest that \SGB was largely successful: in 83.85\% of all requests (161 out of $n$=192), companies responded positively, agreeing to remove ads from the targeted media outlets (\textbf{RQ1}). 
We found that user engagement (likes, replies, retweets) with \SGB's requests and the companies' number of followers had no significant association with whether companies replied to the movement's request.
However, when considering only companies that did reply, we find that pressure rate (\ie, the number of Twitter mentions per minute that the company receives following a request) is correlated with the time between request and answer ($r$=-0.65; $p$<0.001).

Following \SGB's campaigns, we find no evidence that online attention towards the targeted media outlets decreased (\textbf{RQ2}).
Using Twitter and Google Trends data of targeted outlets and of similar outlets that \SGB did not target, we estimate the impact of the campaigns with a Bayesian Structural Time-Series Model~\cite{brodersen2015inferring}.
We find that, in the six months that followed \SGB's campaign, search interest and user engagement towards the targeted outlets did not significantly differ from a synthetic control built with the similar non-targeted outlets.
This finding suggests that, while the removal of ads may have harmed targeted outlets financially, the online attention and engagement towards them did not decrease.

Last, we analyzed whether user interactions with company profiles were impacted in the days, weeks, and the month that followed \SGB's requests and companies' responses (for those that did answer; \textbf{RQ3}). 
We study the evolution of the volume of tweets mentioning companies as well as of the tone of these tweets, as measured by text-derived signals such as Perspective API's toxicity score~\cite{perspective_api} and Sentistrength's \cite{thelwall_sentistrength_2010} positive and negative sentiment scores of a tweet, both obtained using models specifically designed for the Portuguese language.
We find that these metrics experience significant changes following \SGB's requests and companies' answers, but that these changes are transient. All signals systematically return to pre-request levels, regardless of whether companies responded to SGB's request or not.

\xhdr{Implications}
Our results suggest that the Sleeping Giants activism model was successful in getting companies to boycott specific media outlets, which may have harmed their advertisement revenue stream (as claimed by the movement~\cite{oglobo_2021}). 
Yet, we also find evidence for the frailty of the model. Our analyses showed no significant changes to the online attention and engagement going towards targeted companies, nor to the user interactions with companies that did not reply to SGB's requests.
All in all, we hope our findings will serve as a stepping stone for future work that examines online activist movements and that it may inform present and future online activism practices.

\begin{figure}
\centering
\includegraphics[width=0.86\linewidth]{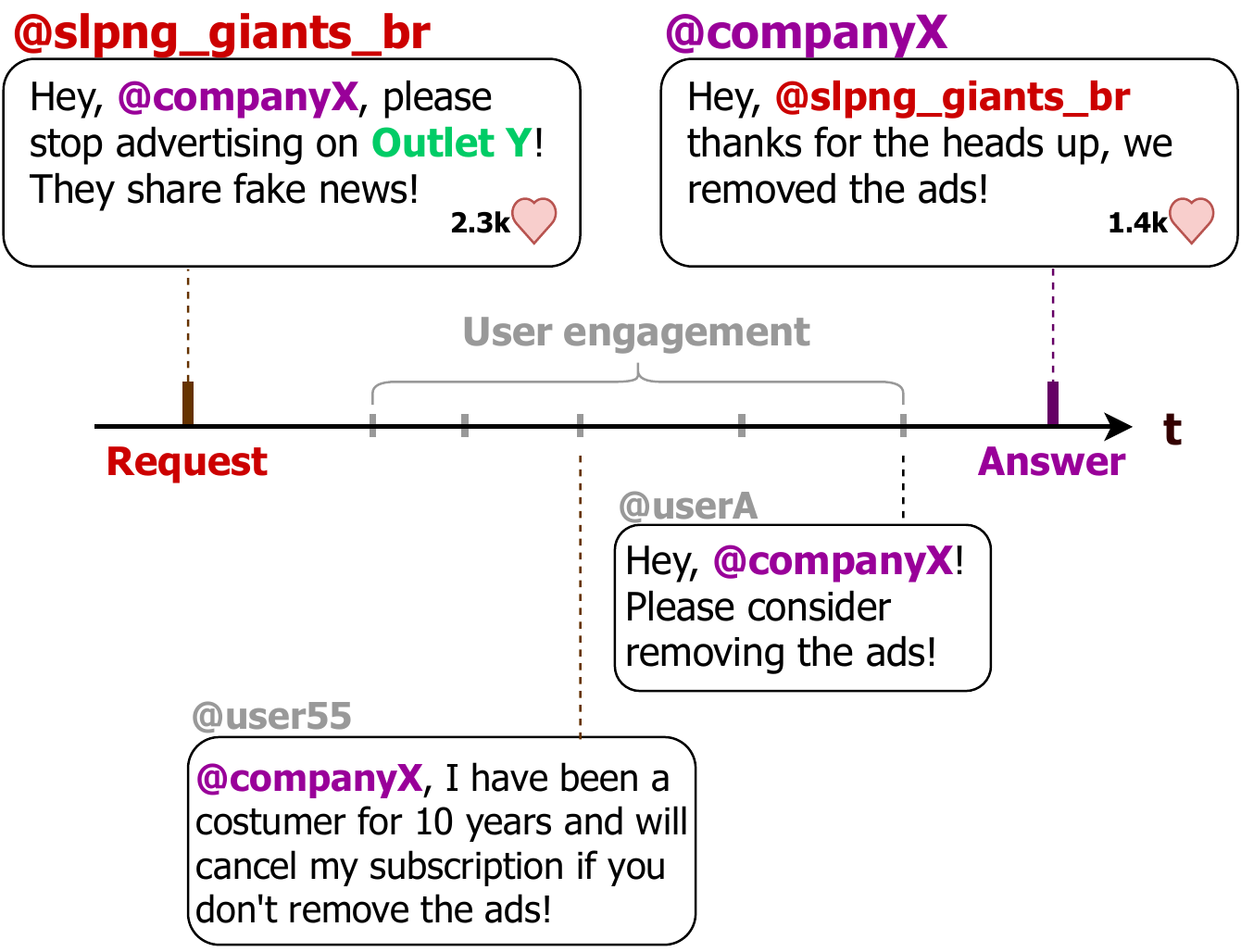}
\caption{Illustration of a \underline{complaint}. \sghandle writes a \underline{request}, which receives some engagement, alerting \textit{@companyX} about ads on \textit{Outlet Y}. Users pressure the company, which eventually provides an \underline{answer}.
The answer is then shared by \SGB.}
\label{fig:actors}
\end{figure}

\section{Background}
\label{sec:bg}

\xhdrNoPeriod{The ``Sleeping Giants Playbook.''}
We briefly describe the activism playbook employed by Sleeping Giants chapters around the globe, including \SGB. First, \SGB chooses a media outlet spreading false or misleading information to target. Second, they find companies advertising on this website (usually through third-party services such as Google ads). 
Third, they issue a ``request,'' \ie, a tweet informing a company that it is advertising on an outlet promoting false information.
This request triggers substantial user engagement, which pressures the company to remove its ads.
Eventually, companies reply to \SGB, often saying they have complied with the request. 
On this occasion, \SGB retweets the company's reply and politely thanks them for their response.
At times, companies either do not reply to \SGB's requests or reply but do not adequately address their request to remove ads. 
In this case, it is common for \SGB to issue further requests.
The ``Sleeping Giants playbook'' (without these additional requests) is illustrated in \Figref{fig:actors}. 
To refer to this sequential process formed by requests and answers involving a given company in a given campaign (\eg, \textit{@CompanyX} and \textit{Outlet Y} in \Figref{fig:actors}) we use the term ``complaint''.

\xhdr{Agents involved} We further describe the key agents present in the previously described process: the targeted media outlets, companies, and users.

\begin{itemize}
\item
\textbf{Targeted media outlets} 
are usually subscription-free media outlets whose revenue stream depends on ads shown to visitors. 
Importantly, these ads are mediated through third-party ad services such as Google Ads, \ie, companies do not explicitly ask to advertise on these websites.

\item 
\textbf{Companies} 
are the direct target of \SGB, acting as a proxy for the movement to harm the revenue stream of the targeted media outlets.
Importantly, companies have the power to block their ads from the targeted media outlets through third-party ad services by placing them in a block list.

\item 
\textbf{Users} 
are responsible for pressuring companies after \SGB makes a request. 
They do this mainly through direct mentions on Twitter, with approaches that vary from polite requests to rude threats to boycott the company. 

\end{itemize}

\section{Related Work}

Since the beginning of the participatory Web, scholars have studied the Internet as a tool for activist movements.
In one of such early works, \citet{bennett2003communicating} suggests that the Internet would go beyond reducing temporal and geographical constraints.
It would facilitate the formation of ``loosely linked, ideologically thin'' issue networks, structures with different strengths and weaknesses compared to activist networks at the time.

These strengths and weaknesses were masterfully illustrated by \citet{tufekci2017twitter}'s comparison between the 1963 March on Washington with the Egyptian Revolution of 2011.
Both protests gathered millions to the streets but were planned in very different time spans: it took months to prepare the March on Washington and only days to mobilize Egyptians to gather in Tahrir Square.
However, while the former movement achieved some of its goals, the latter did not.
According to Tufekci, this illustrates how networked protests (and by extension, networked activism) are remarkably efficient in mobilizing individuals while struggling to challenge the status quo due to their decentralized and non-hierarchical nature.

This fragility has led some to criticize online activism, often called ``slacktivism.'' 
According to \citet{madison2020case}, the term refers to the idea that ``by attempting to carry out political acts online the individual is not participating politically but rather engaging in a form of meaningless, self-serving, and narcissistic acts.''
This angle is explored by \citet{morozov201676}, who argues that slacktivism would be a symptom of a ``naïve belief in the emancipatory nature of online communications.'' 
In his view, while those seeking to promote democracy and progressive values would resort to useless ``slacktivism,'' authoritarian regimes would be efficiently using the internet to stifle dissent and track dissidents.

\begin{table*}[t]
\begin{center}
\small
\caption{We show three tweets among the annotated sample. Original tweets were translated by the authors.}

\begin{tabular}{lp{16cm}}
\toprule
 Label & Tweet\\
\midrule
I-1a  & 
One more company committing itself against the monetization of websites that spread fake news. We can already shop for our buddies at @PetzOficial knowing that the company is socially responsible with its advertising! Thank you for the support \#SleepingGiantsBrasil 
\newline
\small{(Original tweet: https://twitter.com/slpng\_giants\_pt/status/1284276803499175936)}\\

\midrule
I-2a & 
Hello @PetzOficial, how are you? It's great to buy the products our pets need without leaving home, but we found again an advertisement of yours on a website that spreads fake news and hate speech. Plz BLOCK \#SleepingGiantsBrasil 
\newline
\small{(Original tweet: https://twitter.com/slpng\_giants\_pt/status/1276243768111362048)}\\

\midrule
I-2b & Today we complete 33 days without a stance from @SpotifyBR about its advertisements being contributing to websites that spread fake news and hate speech. We would like an answer from the company, and you? WE REITERATE THAT YOU CONFIRM THE BLOCK \#SleepingGiantsBrasil 
\newline
\small{(Original tweet: https://twitter.com/slpng\_giants\_pt/status/1284232070114545665)}\\

\bottomrule
\end{tabular}
\label{tab:annotation-ex}
\end{center}
\end{table*}

This view, however, goes against empirical work showing that slacktivism does increase participation in subsequent civic actions.
For instance,
\citet{lee_slacktivism} shows how the participants partaking in low-cost and low-risk internet activism were significantly more likely to donate money to a related charity supporting the same cause.
Also, in a similar fashion, \citet{vissers2014spill} observe spill-over effects from online to offline protest participation.
More broadly, but also contrary to the notion of feel-good slacktivism, \citet{jackson2020hashtagactivism} have studied how  marginalized groups advance counter-narratives through internet activism. Through an extensive analysis of hashtag campaigns such as \#MeToo and \#BlackLivesMatter, they argue that online activism has shaped journalistic and political narratives in the United States.

In a different line of work, \citet{christensen_slacktivism} examines whether the accusation of ``slacktivism'' is valid. 
He concludes that it is not possible to determine if online activist campaigns are effective, but if anything, they contribute to off-line mobilization.
Although the existing literature suggests that online activism can indeed stimulate off-line mobilization~\cite{lee_slacktivism,vissers2014spill} and increased awareness~\cite{jackson2020hashtagactivism}, there's still little research investigating the concrete impact of online activists on the causes they advocate for. 
In this direction, we highlight the work of \citet{gaffney_iran}, who attempted (with little success) to measure the impact of Twitter activists in the Iranian Election of 2009.

In this context, studying the Sleeping Giants movement can help researchers better understand the impact of online activism. First, it is a recent movement, helping to extend the discussion above to current times, as the participatory Web has considerably evolved in the last decade. And second, unlike other kinds of online activism, the success of campaigns carried out by Sleeping Giants (asking companies to remove ads from specific media outlets) is easily verifiable. 
According to \citet{colli_indirect_strategies}, who studies the  movement as a form of indirect consumer activism, their strategy is a ``concrete and `satisfying’ way to tackle the problem, providing people with a real way to feel that they were having an impact.''

Previous work has discussed the structure and the spread of the Sleeping Giants movement across the world~\cite{braun2019activism}. 
In addition, \citet{li2021beyond} studied Sleeping Giants as a successful case of online activism and investigated the conditions that led to the emergence and popularity of the American chapter, as well as the tactics they employed. In this paper, we expand this literature by quantitatively analyzing the effectiveness of the Sleeping Giants' activism model in Brazil.

\section{Methods}
\label{sec:methods}

We describe the data employed in this paper, how it was annotated, and how we derived meaningful metrics to study the Sleeping Giants activism model.

\subsection{Data and Annotation}

\xhdr{\SGB tweets} 
We collected the entire timeline of \sghandle from May 18, 2020 (creation date), to September 20, 2020, totaling 1560 tweets. We manually selected tweets that are part of a campaign initiated by \SGB, ending up with 3 campaigns and 427 tweets, all of them in Portuguese. 
These tweets contained 501 mentions to companies.
Each campaign was associated with a specific media outlet (\textit{Jornal da Cidade}, \textit{Brasil sem Medo} and \textit{Conexão Política}), and these three were the only online news websites focused by \SGB between May and September 2020. 
In subsequent months (which are outside of the scope of this paper), SGB's campaigns focused on demonetizing and bringing down other kinds of targets such as YouTube channels and influencers. 
Between April and May 2021, after the data collection for this study was completed, SGB conducted a campaign against a news website called \textit{Estudos Nacionais}. 
We do not consider data from this campaign.

\begin{table*}[t]
\centering
\caption{Summary of all signals used in our study, along with a mapping of which signals are used to answer which research question (\cf \Secref{sec:intro}). Note that RQ1, RQ2, and RQ3 are addressed in \Secref{sec:rq1}, \Secref{sec:rq2}, and \Secref{sec:rq3}, respectively. }

\begin{tabular}{llp{11cm}l}
\toprule
Type       & Metric & Description       & Usage       \\ \midrule
\textit{Popularity} & Num. followers  & Number of twitter followers. & \textbf{RQ1} \\    
& Engagement      & Likes + replies + retweets of a tweet. & \textbf{RQ1/RQ2}  \\   
& Search volume   & Google Trends calibrated search volume \cite{west_calibration_2020}. & \textbf{RQ2} \\   \midrule
\textit{Pressure}   & Response time   & Minutes between  SGB's request and the company's answer. & \textbf{RQ1} \\   
& Pressure rate   &    Number of user mentions in a specific time span (\eg, a request and its answer). & \textbf{RQ1/RQ3}     \\     \midrule
\textit{Affect}     & Toxicity        &  Perspective API's \cite{perspective_api} toxicity score. & \textbf{RQ3} \\ 
& Sentiment score &  SentiStrength's \cite{thelwall_sentistrength_2010} positive and negative sentiment scores. & \textbf{RQ3} \\ \bottomrule      
\end{tabular}
\label{tab:metrics}
\end{table*}

\begin{figure}[t]
\centering
\includegraphics[width=\linewidth]{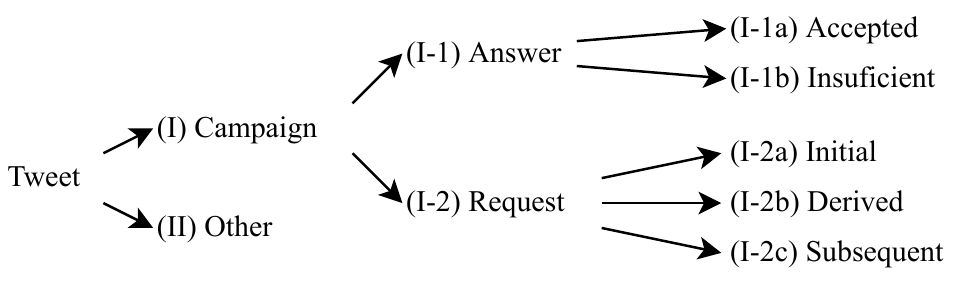}
\caption{Annotation process of \SGB's tweets.}
\label{fig:labels}
\end{figure}

\xhdr{Campaign annotation} 
We manually annotate campaigns with the support of a ``control sheet'' provided to us by the owners of SGB. 
For each campaign, the sheet described which and when companies were targeted and when and how they answered SGB's demands. 
We labeled each mention of a company in a tweet made by SGB in a hierarchical fashion, as illustrated in Fig.~\ref{fig:labels}.
Note that the annotation was done in the mention level, and tweets that mentioned multiple companies were annotated once per company mentioned.

Tweets were annotated as being part of a campaign (I) or not (II), and, if so, as either an ``answer''~(I-1) or a ``request''~(I-2).
Answers are tweets where \SGB retweeted a company's answer to the movement.
They were further classified as either:
``accepted''~(I-1a), \ie, the company made a statement confirming the removal of ads from the venues targeted by \SGB, or 
``insufficient''~(I-1b), when \SGB did not consider the answer satisfactory (and continued to pressure the company to take a stance).
Requests are tweets in which \SGB calls out a company for having ads in one of the targeted media outlets.
They were further classified as either
an ``initial request'' (I-2a), when \SGB targets the company for the first time in a campaign;  
a ``derived request'' (I-2b), when the company did not yet answer and \SGB reinforced the initial request;
or a ``subsequent request'' (I-2c), when the company provided an accepted answer, but advertisements were again seen on the targeted venues.
We illustrate three annotated examples with different labels in \Tabref{tab:annotation-ex}.

\newpage

\xhdr{Targeted companies' mentions and followers}
For each company, we collect the number of followers on Twitter, resorting to the internet archive for deleted profiles.
We also retrieve all tweets mentioning each company targeted by \SGB from 30 days before each initial request to 30 days after the last answer (or initial request for companies that did not answer). 
We ended up with a total of 166 companies, of which 144 answered at least one of \SGB's requests, and 1,079,918 tweets mentioning them. 
As these tweets are used to feed Portuguese-specific models, we use \texttt{langdetect}%
\footnote{https://pypi.org/project/langdetect/}
to certify that the majority of them are indeed in Portuguese.
We find that over 95\% of the tweets were identified as being written in Portuguese. 
A small fraction of the tweets (1.33\%) were identified as being written in English, perhaps due to the usage of terms like "fake news" in short tweets.


\xhdr{Targeted outlets' tweets and search volume} 
Between May 18 and September 20, 2020, three media outlets were targeted by \SGB: \textit{Jornal da Cidade}, \textit{Brasil sem Medo} and \textit{Conexão Política}. 
In the campaigns \SGB issued, respectively, 59, 126, and 7 initial requests. 
For each targeted media outlet, we collect their entire Twitter timeline and the Google Trends search volume for their names 6 months before and after each campaign was initiated. 
To obtain the search volume for each outlet in an absolute scale, we use G-TAB~\cite{west_calibration_2020}, a methodology that allows researchers to overcome Google Trends' limitations.

We additionally use 22 non-targeted outlets%
\footnote{\textit{O Antagonista, República de Curitiba, Folha Política, Gazeta Brasil, Portal Novo Norte, Terra Brasil Notícias, Vista Pátria, Terça Livre, Renova Mídia, Papo TV, Folha Política, Uol, Folha de São Paulo, Correio Braziliense, G1, Estadão, Estado de Minas, R7, Isto É, Metrópoles, Ig, CNN Brasil}} as controls in one of our analyses (\cf \Secref{sec:rq2}), collecting their search volume and Twitter engagement (for the 17 out of these that had a Twitter account).
To obtain these outlets, we used Alexa%
\footnote{https://alexa.com} 
to obtain the top 5 similar sites by audience overlap for each of the targeted media outlets, as well as all news outlets listed among the top 50 most popular sites in Brazil.
For top media outlets, we also included news sites in their 5 most similar sites by audience overlap. 
Finally, we eliminated outlets with an ambiguous name due to the search-volume data.
We detail how these controls are used in \Secref{sec:bsts}. 

\begin{figure}[t]
\centering
\includegraphics[width=\linewidth]{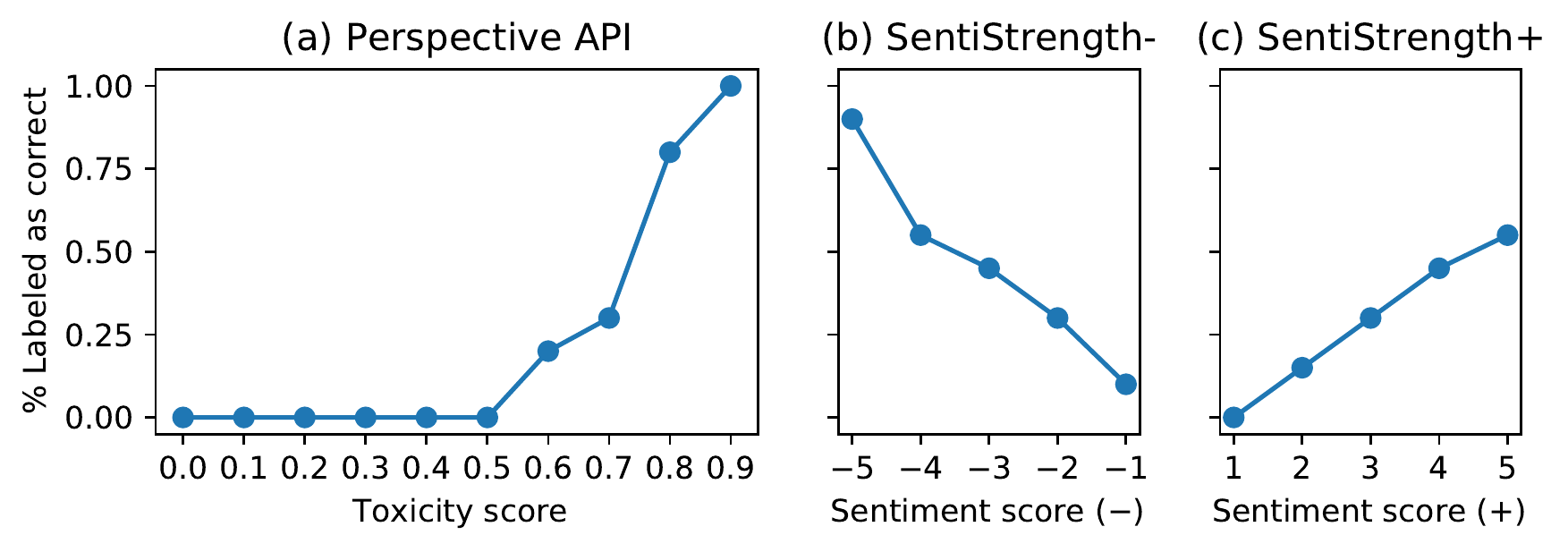}
\caption{Results of our validation of Perspective API and SentiStrength. In the $x$-axis, we show the scores given by either Perspective API or SentiStrength. In the $y$-axis, we show the percentage of tweets with that score that we considered to be indeed toxic/negative/positive in our annotation.}
\label{fig:sent_annotation}
\end{figure}

\subsection{Derived Signals}

From the aforementioned data, we derive popularity\hyp{}, pressure\hyp{}, and affect-related signals. These are briefly summarized in \Tabref{tab:metrics} and described below.

\xhdr{Popularity signals} 
To study popularity, we used three signals: the number of followers that Twitter accounts had, the engagement on tweets (the total number of likes, replies and retweets a tweet received), and the previously described search volume for the targeted (and control) media outlets, .

\xhdr{Pressure-related signals} To study how users pressure companies, and how fast they answer, we obtained two signals from the Twitter data. 
First, we calculated the \textit{response time} for each complaint, \ie how much time it took for companies to answer after \SGB issued an initial request targeting them.
Second, we calculate the \textit{pressure rate} exerted by users. We define the pressure rate as the number of tweets mentioning the company per minute in a specific time span (\eg., the time between \SGB's request and the company's response).

\xhdr{Affect-related signals}
Last, to study how tweets changed following \SGB's campaigns, we employ two text-derived signals.
First, we use Google's Perspective API's \textit{toxicity score}~\cite{perspective_api}. The API consists of machine learning models trained to assess how likely (on a scale from 0 to 1) a text is "rude, disrespectful, or unreasonable, and is likely to make you leave a discussion."
Second, we use SentiStrength's~\cite{thelwall_sentistrength_2010} positive and negative sentiment scores for each tweet. SentiStrength leverages a corpus of short annotated comments to rate the sentiment in sentences in scales from 1 to 5 (for positive sentiment, where 5 is extremely positive) and from -1 to -5 (for negative sentiment, where -5 is extremely negative).
We use versions of both models trained for the Portuguese language.

To validate the results obtained from Perspective API and SentiStrength, we manually annotate a stratified sample of 100 tweets for each score, labeling each of them according to whether they are toxic/positive/negative or not. 
For the Perspective API, we divide the possible scores (\ie, $[0, 1]$) in ten equally sized intervals (\ie, $[0, 0.1)$, $[0.1, 0.2) \ldots$) and sample ten tweets in each interval. For SentiStrength, where the output is always one out of five numbers, we sample 20 tweets for each each output. Results of this validation study are shown in \Figref{fig:sent_annotation}. We find that the signals are largely in accordance with the manual labels, \eg, the chance of a tweet being toxic monotonically increases with their toxicity score.

\begin{table}[t]
\caption{Overview of the complaints initiated by \SGB across the three analyzed campaigns. 
For each kind of request and answer (in the first column), we show the number of tweets (in the second column) and the number of associated complaints. Note that a single tweet may address multiple complaints, \eg, tagging multiple companies at the same time. 
Lastly, in the fourth column, we depict the average engagement (\cf \Secref{sec:bg}) from \SGB tweets associated with each kind of request and answer.}

\begin{center}

\small
\begin{tabular}{lcp{1.75cm}r}
\toprule
\textbf{Label} & \textbf{\#Tweets} & \textbf{\#Complaint} &  \textbf{$\mu_{engagement}$}\\
\midrule
Initial request    &     190 &         192 &          5427.72 \\
Accepted answer      &     159 &         161 (83.85\%) &          3281.26 \\
Insufficient answer  &      11 &          11 (5.73\%) &          4142.64 \\
Subsequent request &       3 &           3 (1.56\%)&          8915.67 \\
Derived request    &      75 &          53 (27.60\%) &          4759.09 \\
\midrule
Total                &     438 &         192 &          4525.65 \\
\bottomrule
\end{tabular}
\label{tab:campaigns}
\end{center}
\end{table}

\subsection{Estimating the impact of SGB's campaigns}
\label{sec:bsts}
To estimate the impact of SGB's campaigns on media outlets we use the \textit{CausalImpact} method proposed by \citet{brodersen2015inferring}. 
Using a Bayesian structural time series model (BSTS), the method constructs a synthetic control from several observed time-series that may be correlated to the time-series that received treatment using data from the pre-treatment period.
Then, in the post-treatment period, the method calculates the causal effect of the treatment based on the distance between the counterfactual (synthetic) time-series and the treated (observed) time-series.
Notably, the model is fully Bayesian, giving flexibility on the ways which posterior inferences can be summarised. 
In other previous work, this methodology has found a wide variety of uses: from estimating the population-level impact of vaccines~\cite{bruhn2017estimating} to measuring the effect of extremist violence on online hate speech~\cite{olteanu2018effect}.
In the work at hand, in \Secref{sec:rq2}, we use the search volume and engagement time-series from the media outlets obtained from Alexa as controls (used to build the synthetic control in \textit{CausalImpact}) and the time-series from targeted outlets as the treated time-series. 
We run a separate analysis for each of the three targeted outlets.

\begin{figure}[t]
\centering
\includegraphics[width=\linewidth]{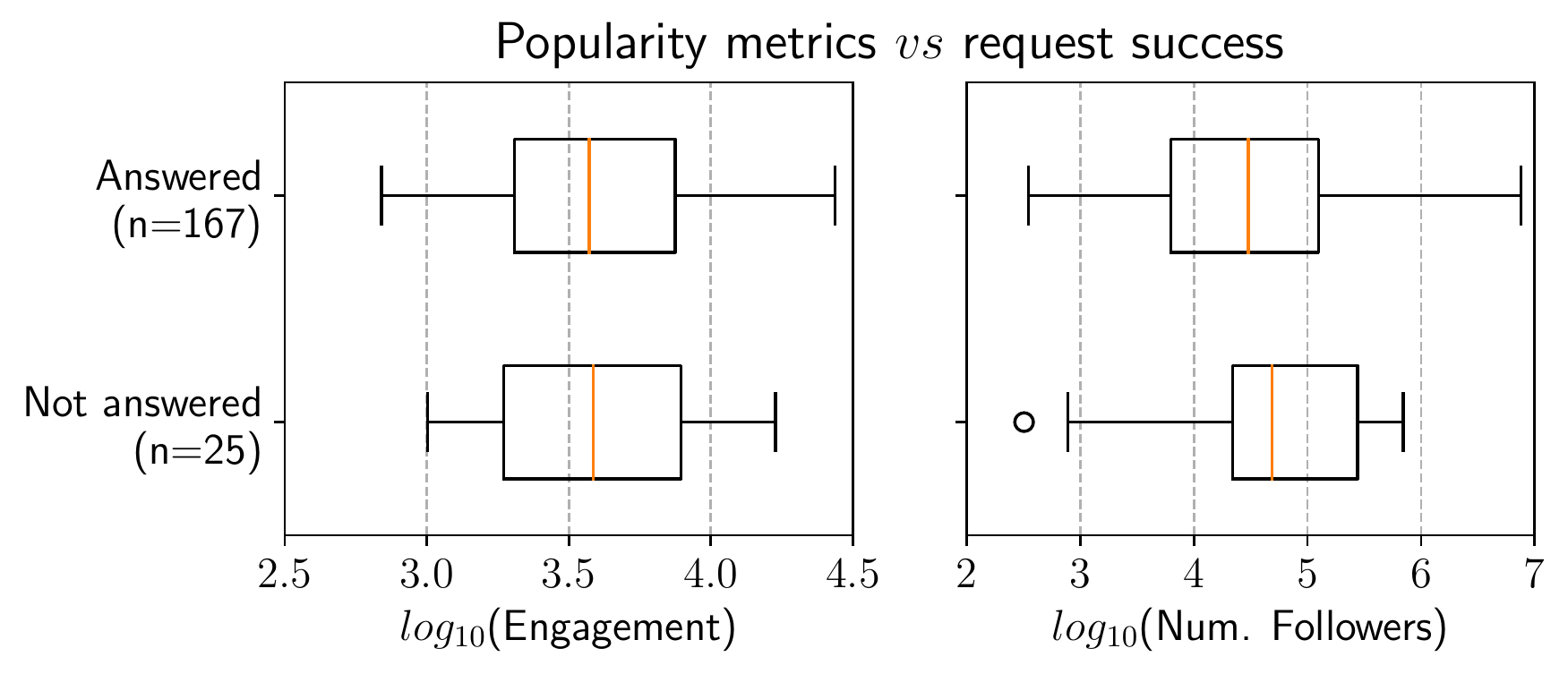}
\caption{Box-plot showing the $\log$-transformed distribution of engagement (left) and number of followers (right) for answered and not answered requests. 
Box boundaries mark quartiles, the middle bar the median, and whiskers the 5th and 95th percentiles.}
\label{fig:when1}
\end{figure}

\section{Results}
\subsection{RQ1: Measuring the success of SGB's requests}
\label{sec:rq1}

We begin investigating the success of \SGB's requests, \ie, asking companies them to remove ads from a specific media outlet. We analyze these requests (and eventual answers) at the complaint level. A complaint (as discussed in \Secref{sec:bg}) encompasses all requests and answers associated with a company in a given campaign.

\begin{figure*}[t]
\centering
\includegraphics[width=\linewidth]{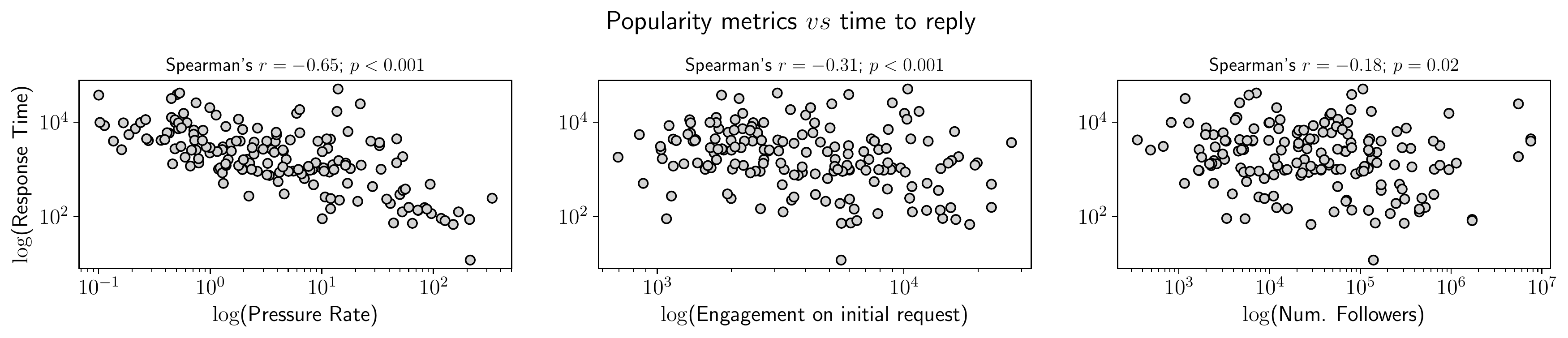}
\caption{We show the relationship between three popularity metrics (in the $x$-axis) and the time to reply (in the $y$-axis). On top of each plot, we depict the Spearman's correlation coefficient for each pair of variables. Note that all axes are scaled logarithmically, and that only the $y$-axes are shared across plots.}
\label{fig:fast2}
\end{figure*}

\begin{figure}
\centering
\includegraphics[width=\linewidth]{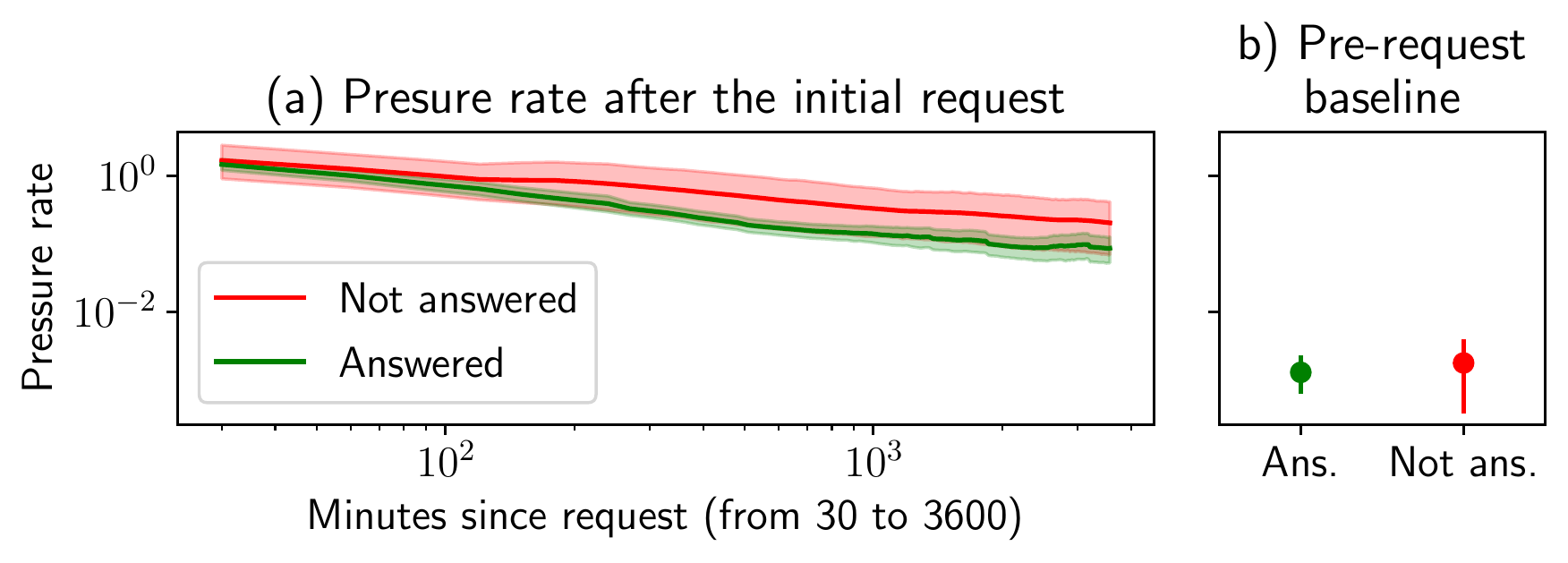}
\caption{
Comparison between pressure rate for answered and not answered \SGB requests.
In \textit{a}, we compare the pressure rate for requests that were not answered with those that were answered after $t$ minutes ($x$-axis). In \textit{b}, we compare the pressure rate for a 5-day period before the requests were made. Shaded areas (in \textit{a}) and error bars (in \textit{b}) depict bootstrapped 95\% CIs.
}
\vspace{-4mm}
\label{fig:when2}
\end{figure}

\xhdrNoPeriod{Are requests answered?} 
Overall, \SGB's requests were largely successful with 83.85\% of all 192 initial requests being (eventually) acceptably answered.
Some of these answers were considered insufficient and not accepted by \SGB, but even then all companies in our dataset made a positive statement, confirming the removal of their advertisements, committing themselves to do so, or just stating that they are against the spread of misinformation.
We depict statistics related to \SGB's complaints in \Tabref{tab:campaigns}. 
We note that 27.6\% of the times, initial requests were succeeded by derived requests (\cf \Secref{sec:bg}).
Also, we find that requests receive on average more engagement than answers: the average engagement in the tweets labelled as ``initial request'' tweets was 5427.7 (\vs 3281.3 for those labelled as ``accepted answer'').

\xhdrNoPeriod{What factors are associated to requests being answered?} 
Next, we investigate the relationship between popularity metrics and whether requests were answered or not. 
We start by comparing the number of followers each company had and the engagement in the initial request, as shown in \Figref{fig:when1}. Using a Mann-Whitney $U$ test, we are not able to reject the null hypothesis that both distributions (answered \vs not answered) are equal
(for \emph{engagement}, $U$=2087, $p$=0.5; for \emph{company followers}, $U$=1746, $p$=0.09).

Another interesting popularity metric to analyze is the \emph{pressure rate}. However, here we do not have a ``response time'' for the companies that have never replied (\cf \Secref{sec:methods}).
To address this, we conduct an additional study where, for each minute $t$ after the initial request was made, we compare the pressure rate up to $t$ between requests that have never been answered and requests that were eventually answered after minute $t$. We show the results for this analysis in \Figref{fig:when2}~(a). Again, we find little difference between the pressure exerted on companies that have never answered the requests and those that eventually did. 
Although the former has a slightly higher pressure rate, point-wise Mann-Whitney U tests consistently fail to reject the null hypothesis that the two distributions are different ($p > 0.05$). As a sanity check, we also analyze the pressure rate in the 5 days before the request, as shown in \Figref{fig:when2}~(b), and find no significant difference between both distributions ($U$=1578, $p$=0.3).

\begin{figure}[t]
\centering
\includegraphics[width=\linewidth]{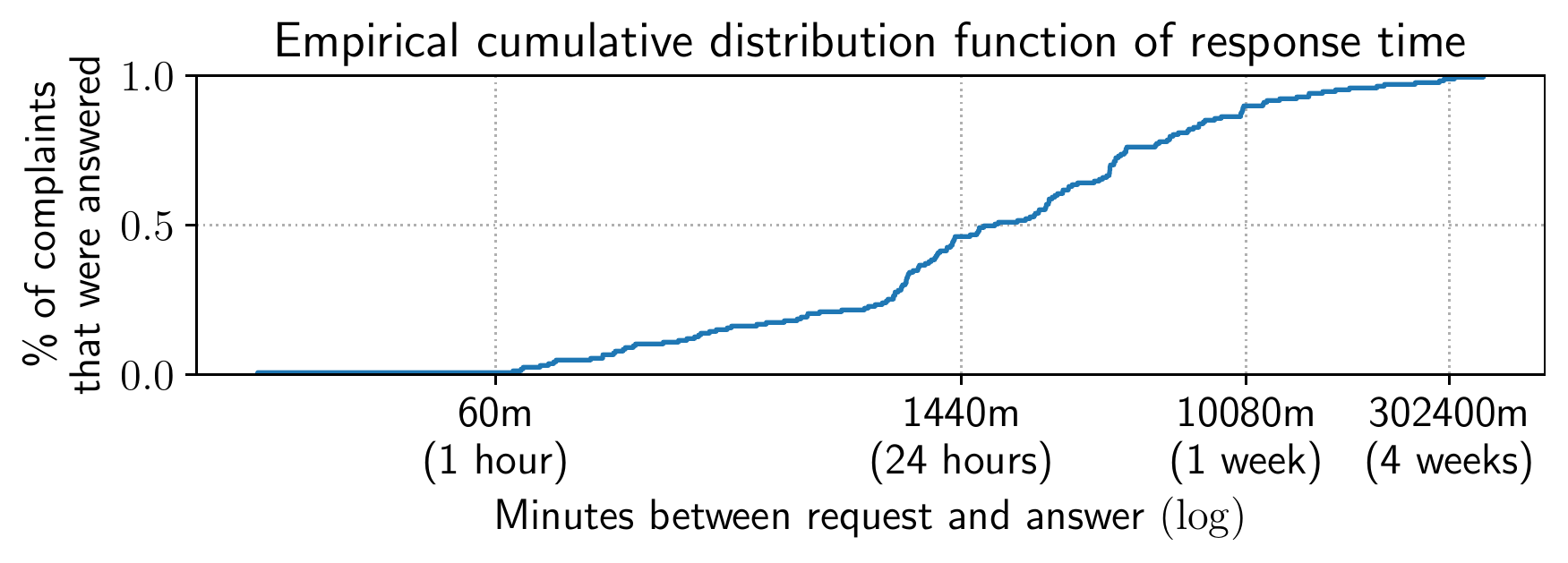}
\caption{Empirical CDF of \SGB requests' response time (in minutes). The $x$-axis is logarithmically scaled.}
\label{fig:fast1}

\end{figure}

\xhdrNoPeriod{How fast are requests answered?} 
To further explore the success of \SGB's requests, we conduct a subsequent analysis considering only requests that were answered (n=167) and studying the time companies take to respond to them. Note that here we consider the first time a company answers to a request, even when \SGB does not accept the answer (\cf, \Secref{sec:bg}).
\Figref{fig:fast1} shows how much time it took for each initial request to be answered. 
We observe that 89.82\% of the initial requests are answered within a week, 46.10\% within 24 hours, and 22.16\% on the same day.

\xhdrNoPeriod{What factors are correlated to the speed of answers?} 
Lastly, we analyze the relationship between popularity metrics and the speed at which companies answer, \ie, response time. Since the distributions of both the response time and the popularity metrics are heavy-tailed, we consider the $\log_{10}$ of these values.

In \Figref{fig:fast2} we portray the relationship between the response time and the three popularity metrics studied (pressure rate, engagement in the request tweet, and number of followers). 
Notice that here we find statistically significant correlations between all three of these variables. This is in contrast to what we observed when studying the success of the requests.
Although we did not find evidence that users tagging the company or liking \SGB's request tweet led to companies answering more often, here we find that these factors are correlated with \textit{how fast} companies answer them (Spearman's $r$=-0.65 for pressure rate ($p$<0.001);
$r$=-0.31 for engagement ($p$<0.001).
The number of followers that tagged companies had at the request time is also significantly correlated with response time ($p$=0.02), but the correlation is weaker ($r$=-0.18).

\begin{figure*}
    \centering
    \includegraphics[width=\linewidth]{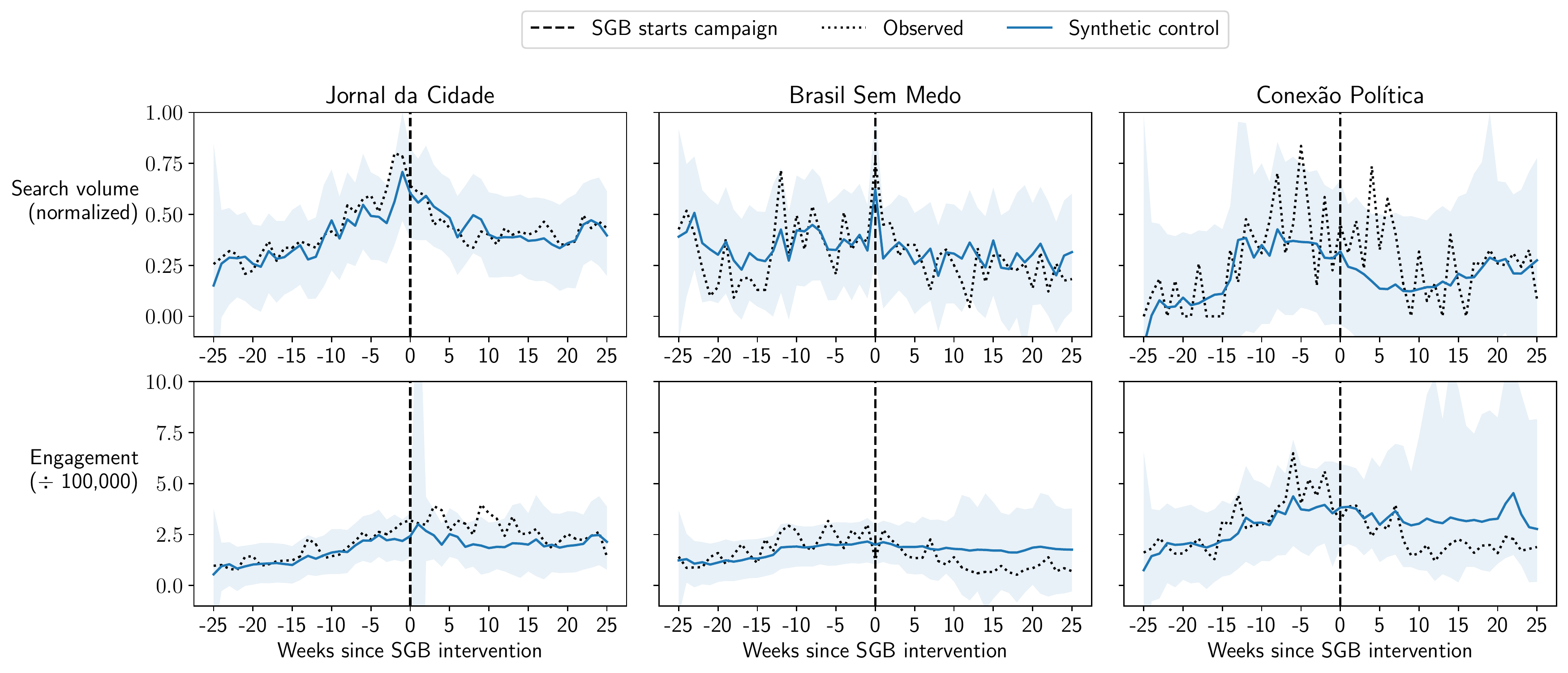}
    \caption{We depict the search volume (top row) and the engagement (bottom row) for the targeted outlets (in dotted black, one per column) and for the synthetic controls created using the \textit{CausalImpact} method (in blue). The colored blue area represents the 95\% credible interval for the Bayesian method --- 95\%  of probability density lies within that interval.}
    \label{fig:eng}
\end{figure*}

To better understand the joint effect of these three variables in the time to reply, we perform a regression analysis considering the log-transformed response time as the response variable, and the pressure rate, engagement on the initial request tweet, and number of followers as the explanatory variables (all of them also log-transformed). 
More precisely, our model is described by:
\begin{equation*}
\begin{split}
\ln(\text{Time to reply}) = \beta_0 &+ \beta_1 \ln(\text{Pressure rate}) 
\\ &+ \beta_2 \ln(\text{Engagement}) 
\\ &+ \beta_3 \ln(\text{Num. Followers}).
\end{split}
\end{equation*}
Since both response and explanatory variables are log-transformed (with the natural logarithm), we can interpret the coefficients as \textit{the percent increase in the response variable for every 1\% increase in the explanatory variable.}

Results for this regression analysis are shown in \Tabref{tab:reg1}.
We find that all coefficients are statistically significant (with $p<0.001$) but that only the pressure rate has a negative coefficient. 
This coefficient, $\beta_1 = -1.0$, can be interpreted as follows: for every $1\%$ increase in the pressure rate, there is a 1\% decrease in the time to reply.
The two other variables, on the contrary, are correlated with an \textit{increase} in the time to reply when the pressure rate is taken into account ($\beta_2 = 0.88$; $\beta_3 = 0.22$).
A hypothesis that explains this result is that the pressure rate is a mediator for the other two variables, \ie, a high number of followers and engagement on the request would lead to a higher pressure rate.

\begin{table}[]
\centering
\caption{Regression results analyzing the (log-transformed) time to reply as a function of popularity metrics. *** indicates that p-values were smaller than 0.001. $R^2=0.64$.}
\small

\begin{tabular}{lcccccc}
\toprule
& \textbf{Coef} & \textbf{Std err}  \\
\midrule
\textbf{Intercept}                &      -4.9***   &        1.295      \\
\textbf{ln(Pressure rate)}   &      -1.0***   &        0.071       \\
\textbf{ln(Engagement)}    &       0.88***   &        0.152       \\
\textbf{ln(Num. Followers)} &       0.22***   &        0.050         \\ \bottomrule
\end{tabular}
\label{tab:reg1}
\end{table}

\xhdr{Takeaways} \SGB requests are successful: most companies eventually reply to their requests. We found no evidence that popularity metrics related to the requests influence whether companies answer requests. Still, we see a significant correlation between these popularity metrics and the time it takes companies to respond.

\subsection{RQ2: Measuring changes to the targeted media outlets}
\label{sec:rq2}

We now turn to analyze the effect of \SGB's campaigns for all three media outlets they targeted: \textit{Jornal da Cidade}, \textit{Brasil sem Medo} and \textit{Conexão Política}. 
To do so, we leverage two metrics: the engagement in tweets made by the official accounts of the media outlet and the search volume towards them.
As discussed in \Secref{sec:methods}, we use the \textit{CausalImpact} method proposed by \citet{brodersen2015inferring}, contrasting the evolution of engagement and search volume of these outlets with the evolution of a synthetic control. 

More specifically, for each of the three targeted outlets, we consider the 25 weeks before the start of SGB's campaign and create a synthetic control using the signals of non-targeted outlets. Then, for the 25 weeks following SGB's campaign, we estimate its effect on the targeted outlet by contrasting the observed engagement and search volume with predictions from the control time series.
To facilitate comparison across different targeted outlets, we normalize the search volume so that the highest observed value equals 1.

We depict the results of this analysis in  \Figref{fig:eng} for search volume (top) and engagement (bottom). 
For each signal and each targeted outlet, we also calculate the cumulative relative effect and the corresponding posterior tail-area probability (\textit{pp}). 
An explanation of these summary statistics as well as their values for each $\langle$signal, outlet$\rangle$ pair are shown in \Tabref{tab:ci}.

Overall, both the image and the posterior tail-area probability (\textit{pp}) suggest that there is little difference between the targeted outlets and the synthetic controls, implying that SGB's campaigns did not have an effect in reducing online attention going towards these outlets. 
Note that, although the cumulative relative effects are at times negative (\eg,  -36.12\% for \textit{Brasil Sem Medo}, engagement), the \textit{pp}s indicates that this effect is not statistically significant and so cannot be meaningfully interpreted.

We stress that the usage of synthetic controls in this analysis is essential as the three \SGB campaigns we study happened during the first wave of the COVID\hyp{}19 pandemic in Brazil. The pandemic has significantly increased internet usage across the world~\cite{feldmann2021implications}, and we indeed observe increases in search volume and engagement for the targeted outlets (as well as for many non-targeted ones) between May and July. This period coincides with the rise in cases and the adoption of non-pharmaceutical interventions in Brazil.
In that context, when using \textit{CausalImpact}, we compare the targeted outlets with synthetic controls built with real data from other outlets that experienced a similar shock. 
Thus, \textit{CausalImpact} allows us to obtain an estimate of the effect even with the presence of a confounder (COVID\hyp{}19), which would not be possible with other methods that rely solely on the past values for the time-series of interest (\eg, interrupted time series analysis~\cite{bernal2017interrupted}).

\xhdr{Takeaways}
Overall, our analysis failed to identify a substantial change in the online attention received by outlets targeted by \SGB's campaigns (when contrasted with synthetic controls built with non-targeted outlets).

\begin{table}[]
\centering
\caption{
Summary statistics of the \textit{CausalImpact} method.
For each signal and each venue, we depict the posterior tail-area probability (\textit{pp}) and the cumulative relative effect.\\
\textit{Posterior tail-area probability:} the probability of observing the actual observed data given the model trained.\\
\textit{Cumulative relative effect:}
the relative difference between the synthetic controls and the observed values summed across the 25 weeks following the start of \SGB's campaigns.
}

\begin{tabular}{llrr}
\toprule
Signal & Venue & \textit{pp} & Cum. rel. effect \\
\midrule
Engagement & Jornal da Cidade &  0.09 &     28.69\% \\
           & Brasil Sem Medo &  0.09 &    -36.12\% \\
           & Conexão Política &  0.06 &    -31.57\% \\ \midrule
Popularity & Jornal da Cidade &  0.48 &     -0.51\% \\
           & Brasil Sem Medo &  0.26 &    -12.88\% \\
           & Conexão Política &  0.21 &      36.4\% \\
\bottomrule
\end{tabular}
\label{tab:ci}
\end{table}

\subsection{RQ3: Measuring changes to the targeted companies}
\label{sec:rq3}

Lastly, we investigate the changes in the way social media users interact with companies on Twitter in the weeks following \SGB's request and companies' answers. 

We analyze changes in the pressure rate and text-derived signals (toxicity, positive sentiment and negative sentiment; \cf \Secref{sec:methods}) in tweets mentioning the companies in two different scenarios: 
1) after companies received \SGB's request; 
and 2) after companies answered \SGB's request (for those who answered).
In both cases, we measure the intensity of these signals in four different intervals following each event,\footnote{up to 12h, from 12-24h, from 24-72h, and from 72h to 1 month} and compare it with a baseline calculated in the five days before the request was made, as in \Secref{sec:rq1}.
Note that, for scenario \textit{1}, we consider only companies that did not respond up to the end of that interval (\eg, for 24-72h, we consider only those that never replied and those that replied after 72h).
This way, scenario \textit{1} captures changes succeeding requests with no interference from answers, which are analyzed in scenario \textit{2}.

We make two remarks regarding this analysis.
First, we measure effects on \textit{company-level}, such that all companies have the same weight, even though some receive more mentions than others.
To do so, we calculate, for each company, the average value of each signal in each interval. Then, we report the bootstrapped average value across all companies.
Second, there were twenty-five companies targeted more than once by \SGB.
To not confound changes in the signals with mentions coming from posterior requests, we consider only the first request made for each company and exclude those targeted again by \SGB within the timespan used in this analysis.
For a similar reason, we consider only the last answer given by companies (recall that some provided a new answer if \SGB did not accept the first one).

\begin{figure}[t]
    \centering
    \includegraphics[width=\linewidth]{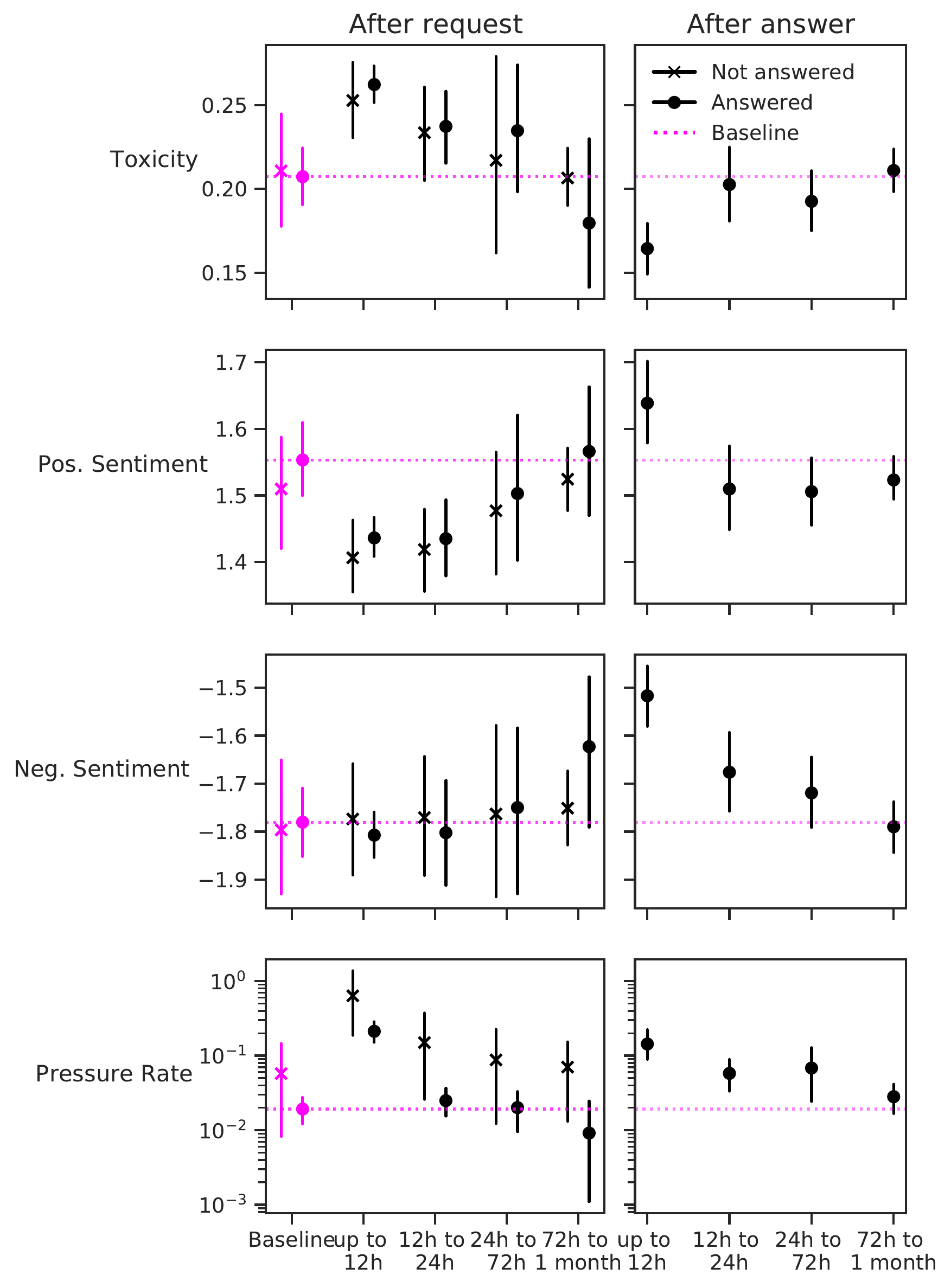}
    \caption{
    The evolution of mention-related signals post-requests and post-answers.
    Each row represents the evolution of a different signal either after the request (in the first column) or the reply (in the second column). 
    For the first column (after the request) we show both companies that did reply ($\circ$) and companies that did not reply ($\times$). Pre-request baselines are shown in magenta in the first column, and, additionally, as a horizontal dotted line for the companies that did reply.
Note the that the $y$-axis of the last row is in logarithmic scale.}
    \label{fig:mentions}
\end{figure}

We show the post\hyp{}request and post\hyp{}answer changes for the previously mentioned signals in \Figref{fig:mentions}. 
For toxicity (1st row) and pressure rate (4th row), we see a sharp increase in the 12 hours following the request (1st column), which gradually decreases as time passes for companies that did and did not reply.
After companies answer (2nd column), we observe a transient \emph{decrease} in toxicity and a transient \textit{increase} in the pressure rate, both of which eventually return to values similar to the pre-request baseline.

Results are similar for negative and positive sentiment signals (2nd and 3rd rows). We observe a sharp decrease in the positive sentiment  following the request and a slight reduction in the negative sentiment, both eventually returning to pre-request levels. After the answer, we note a transient sharp increase in both signals. 
Two exceptions here are the negative sentiment for companies that did not reply, which remains largely unchanged.

\xhdr{Takeaways} 
Overall, this analysis suggests that \SGB's requests and companies answers' change how users interact with the targeted companies. For instance, tweets mentioning the company are more toxic and less positive following the initial request. However, we find that these changes are transient and that interactions eventually become similar to before the initial request. This happens even for companies that did not reply to \SGB's requests.

\section{Discussion}
\label{sec:discussion}
This work quantitatively characterizes Sleeping Giants' ``activism model'' through an in-depth case study of its most popular chapter: \sghandle (\SGB).  By studying a large and impactful activist movement in Brazil, we broaden the scope of the literature studying online activism, which often emphasizes countries in the Global North.
Overall, our work reveals a dichotomy in \SGB's effectiveness, revealing strengths and weaknesses of online activism similar to those qualitatively discussed by previous work~\cite{bennett2003communicating, tufekci2017twitter}. 
On the one hand, our analysis showed that \SGB's requests are largely successful: companies frequently respond to \SGB, and those that answer the quickest are usually the ones that have experienced substantial user pressure (\textbf{RQ1}).
On the other hand,  we did not identify significant changes in the online attention going towards media outlets targeted by \SGB, as they progressed similarly to  the synthetic control we built with non-targeted outlets (\textbf{RQ2}).
Further, we found that the changes in how social media users talk with companies on Twitter are transient and return to normality after a few weeks (\textbf{RQ3}).

\xhdr{Limitations} 
We focus on a specific (and highly successful) chapter of Sleeping Giants (\sghandle).
As such, some of our analyses are limited to relatively small sample sizes. This is particularly tricky in \textbf{RQ1}, where we study \textit{what factors} influence companies to respond to Sleeping Giants' requests. Since most companies have replied (83.85\%, $n$=192), we are left with inconclusive findings on how user pressure and company size impact their likelihood to comply with \SGB's demands. Similarly, in \textbf{RQ2} our analysis is restricted to the only three media outlets that SGB targeted during 2020, and hence the generalizability of our results is not warranted.

Additionally, online activist movements can have spill-over effects \cite{lee_slacktivism, vissers2014spill, jackson2020hashtagactivism} in the real world, but we do not assess those. Therefore, despite the negative findings in \textbf{RQ2} and \textbf{RQ3}, the \SGB movement may have helped increase the collective awareness on online misinformation or inhibited the rise of new outlets sharing false content.
The same applies to possible backlashes to the movement. 
The idea of activism ``backlash'' has been explored by previous research showing that negative stereotypes around activists hinder social change \cite{bashir_activism_stereotype} and that reactions caused by online activism can nullify the progress it brings~\cite{ghobadi_mass_approach}. 

\xhdr{Future work}
An interesting direction for future work is to scale up our analyses.
This could be done across other Sleeping Giants campaigns, chapters in other countries, or even other initiatives that leverage a similar strategy, \eg, U.S.-based \textit{Grab Your Wallet} or U.K.\hyp{}based \textit{Stop Funding Hate}~\cite{colli_indirect_strategies}.
After the period considered in this paper, \SGB has changed its targets to include YouTube channels (\eg, \textit{Terça Livre}) and influencers (\eg, Olavo de Carvalho, a right-wing polemicist).
The methodology we employed could arguably be extended to study these other targets.

Another fruitful direction is to explore the polarization created around the movement and how it connects to its success. 
Since \SGB targeted mainly right-wing outlets, criticism of bias and an anti\hyp{}\sghandle movement also rose on Twitter.
This movement, "Awake Giants," tried to convince  companies \textit{not} to remove ads from the media outlets targeted by \SGB's.
In a recent report, \citet{ufrj} studied the polarized network of interactions between pro\hyp{} and anti\hyp{}\sghandle profiles on Twitter, observing a high level of automated tweets coming from the anti\hyp{}\sghandle profiles.
Further analysis on the content of tweets and on the role of bots in the debate sparkled by \SGB could yield interesting insights on how internet activism and its backlash interact.

\newpage
{ 
\bibliography{refs}

\begin{thebibliography}{30}
\providecommand{\natexlab}[1]{#1}
\providecommand{\url}[1]{\texttt{#1}}
\providecommand{\urlprefix}{URL }
\expandafter\ifx\csname urlstyle\endcsname\relax
  \providecommand{\doi}[1]{doi:\discretionary{}{}{}#1}\else
  \providecommand{\doi}{doi:\discretionary{}{}{}\begingroup
  \urlstyle{rm}\Url}\fi

\bibitem[{Alexandre(2020)}]{gq2020}
Alexandre, E. 2020.
\newblock Sleeping Giants Brasil: o perfil que quer acabar com o financiamento
  de sites que espalham fake news no país.
\newblock In \emph{GQ}.

\bibitem[{Anderson and Rainie(2017)}]{anderson2017future}
Anderson, J.; and Rainie, L. 2017.
\newblock The future of truth and misinformation online.
\newblock \emph{Pew Research Center} 19.

\bibitem[{Bashir et~al.(2013)Bashir, Lockwood, Chasteen, Nadolny, and
  Noyes}]{bashir_activism_stereotype}
Bashir, N.~Y.; Lockwood, P.; Chasteen, A.~L.; Nadolny, D.; and Noyes, I. 2013.
\newblock The ironic impact of activists: Negative stereotypes reduce social
  change influence.
\newblock \emph{European Journal of Social Psychology} .

\bibitem[{Bennett(2003)}]{bennett2003communicating}
Bennett, W. 2003.
\newblock Communicating global activism.
\newblock \emph{Information, Communication \& Society} .

\bibitem[{Bergamo and Finotti(2020)}]{monica_bergamo_2020}
Bergamo, M.; and Finotti, I. 2020.
\newblock Sleeping Giants é formado por casal de 22 anos do interior do
  Paraná.
\newblock In \emph{Folha de São Paulo}.

\bibitem[{Bernal, Cummins, and Gasparrini(2017)}]{bernal2017interrupted}
Bernal, J.~L.; Cummins, S.; and Gasparrini, A. 2017.
\newblock Interrupted time series regression for the evaluation of public
  health interventions: a tutorial.
\newblock \emph{International journal of epidemiology} 46(1): 348--355.

\bibitem[{Braun, Coakley, and West(2019{\natexlab{a}})}]{braun_activism_2019}
Braun, J.; Coakley, J.; and West, E. 2019{\natexlab{a}}.
\newblock Activism, {Advertising}, and {Far}-{Right} {Media}: {The} {Case} of
  {Sleeping} {Giants}.
\newblock \emph{Media and Communication} .

\bibitem[{Braun, Coakley, and West(2019{\natexlab{b}})}]{braun2019activism}
Braun, J.~A.; Coakley, J.~D.; and West, E. 2019{\natexlab{b}}.
\newblock Activism, Advertising, and Far-Right Media: The Case of Sleeping
  Giants.
\newblock \emph{Media and Communication} .

\bibitem[{Brodersen et~al.(2015)Brodersen, Gallusser, Koehler, Remy, and
  Scott}]{brodersen2015inferring}
Brodersen, K.~H.; Gallusser, F.; Koehler, J.; Remy, N.; and Scott, S.~L. 2015.
\newblock Inferring causal impact using Bayesian structural time-series models.
\newblock \emph{The Annals of Applied Statistics} 9(1): 247--274.

\bibitem[{Bruhn et~al.(2017)Bruhn, Hetterich, Schuck-Paim, K{\"u}r{\"u}m,
  Taylor, Lustig, Shapiro, Warren, Simonsen, and
  Weinberger}]{bruhn2017estimating}
Bruhn, C.~A.; Hetterich, S.; Schuck-Paim, C.; K{\"u}r{\"u}m, E.; Taylor, R.~J.;
  Lustig, R.; Shapiro, E.~D.; Warren, J.~L.; Simonsen, L.; and Weinberger,
  D.~M. 2017.
\newblock Estimating the population-level impact of vaccines using synthetic
  controls.
\newblock \emph{Proceedings of the National Academy of Sciences} .

\bibitem[{Christensen(2011)}]{christensen_slacktivism}
Christensen, H.~S. 2011.
\newblock Political activities on the Internet: Slacktivism or political
  participation by other means?
\newblock \emph{First Monday} .

\bibitem[{Colli(2020)}]{colli_indirect_strategies}
Colli, F. 2020.
\newblock Indirect consumer activism and politics in the market.
\newblock \emph{Social Movement Studies} .

\bibitem[{Faris et~al.(2017)Faris, Roberts, Etling, Bourassa, Zuckerman, and
  Benkler}]{faris_partisanship_2017}
Faris, R.; Roberts, H.; Etling, B.; Bourassa, N.; Zuckerman, E.; and Benkler,
  Y. 2017.
\newblock Partisanship, {Propaganda}, and {Disinformation}: {Online} {Media}
  and the 2016 {U}.{S}. {Presidential} {Election}.
\newblock Technical report, Social Science Research Network.

\bibitem[{{Feldmann et al.}(2021)}]{feldmann2021implications}
{Feldmann et al.}, A. 2021.
\newblock Implications of the COVID-19 Pandemic on the Internet Traffic.
\newblock In \emph{Broadband Coverage in Germany; 15th ITG-Symposium}.

\bibitem[{Gaffney(2010)}]{gaffney_iran}
Gaffney, D. 2010.
\newblock iranElection: Quantifying online activism.
\newblock In \emph{In Proceedings of the Web Science Conference}.

\bibitem[{Ghobadi and Clegg(2015)}]{ghobadi_mass_approach}
Ghobadi, S.; and Clegg, S. 2015.
\newblock A Critical Mass Approach to Online Activism: "These days will never
  be forgotten …".
\newblock \emph{Information and Organization} .

\bibitem[{Jackson, Bailey, and Welles(2020)}]{jackson2020hashtagactivism}
Jackson, S.~J.; Bailey, M.; and Welles, B.~F. 2020.
\newblock \emph{\#HashtagActivism: Networks of race and gender justice}.
\newblock MIT Press.

\bibitem[{Jigsaw(2021)}]{perspective_api}
Jigsaw. 2021.
\newblock Perspective API.
\newblock \url{https://www.perspectiveapi.com}.

\bibitem[{Lee and Hsieh(2013)}]{lee_slacktivism}
Lee, Y.-H.; and Hsieh, G. 2013.
\newblock Does Slacktivism Hurt Activism? The Effects of Moral Balancing and
  Consistency in Online Activism.
\newblock In \emph{Proceedings of the SIGCHI conference on human factors in
  computing systems}.

\bibitem[{Li, Bernard, and Luczak-Roesch(2021)}]{li2021beyond}
Li, Y.; Bernard, J.-G.; and Luczak-Roesch, M. 2021.
\newblock Beyond Clicktivism: What Makes Digitally Native Activism Effective?
  An Exploration of the Sleeping Giants Movement.
\newblock \emph{Social Media+ Society} .

\bibitem[{Madison and Klang(2020)}]{madison2020case}
Madison, N.; and Klang, M. 2020.
\newblock The Case for Digital Activism: Refuting the Fallacies of Slacktivism.
\newblock \emph{Journal of Digital Social Research} 2(2): 28--47.

\bibitem[{Morozov(2016)}]{morozov201676}
Morozov, E. 2016.
\newblock \emph{The Net Delusion: How Not to Liberate the World}.
\newblock Columbia University Press.

\bibitem[{Noia(2021)}]{oglobo_2021}
Noia, J. 2021.
\newblock Portais que divulgam fake news perderam R\$ 5,4 milhões em
  publicidades em dez meses, segundo Sleeping Giants Brasil.
\newblock In \emph{O Globo}.

\bibitem[{Olteanu et~al.(2018)Olteanu, Castillo, Boy, and
  Varshney}]{olteanu2018effect}
Olteanu, A.; Castillo, C.; Boy, J.; and Varshney, K. 2018.
\newblock The effect of extremist violence on hateful speech online.
\newblock In \emph{Proceedings of the international AAAI conference on web and
  social media}, volume~12.

\bibitem[{Rollsing(2020)}]{sg_zerohora_2020}
Rollsing, C. 2020.
\newblock Os detalhes da decisão judicial que determinou a identificação dos
  responsáveis pelo Sleeping Giants.
\newblock In \emph{Zero Hora}.

\bibitem[{{Santini et. al}(2022)}]{ufrj}
{Santini et. al}, R.~M. 2022.
\newblock {Relatório da Rede Sleeping Giants Brasil no Twitter}.
\newblock \url{https://bit.ly/3gNlcXK}.

\bibitem[{Thelwall et~al.(2010)Thelwall, Buckley, Paltoglou, Cai, and
  Kappas}]{thelwall_sentistrength_2010}
Thelwall, M.; Buckley, K.; Paltoglou, G.; Cai, D.; and Kappas, A. 2010.
\newblock Sentiment Strength Detection in Short Informal Text.
\newblock \emph{J. Am. Soc. Inf. Sci. Technol.} .

\bibitem[{Tufekci(2017)}]{tufekci2017twitter}
Tufekci, Z. 2017.
\newblock \emph{Twitter and tear gas: The power and fragility of networked
  protest}.
\newblock Yale University Press.

\bibitem[{Vissers and Stolle(2014)}]{vissers2014spill}
Vissers, S.; and Stolle, D. 2014.
\newblock Spill-over effects between Facebook and on/offline political
  participation? Evidence from a two-wave panel study.
\newblock \emph{Journal of Information Technology \& Politics} 11(3): 259--275.

\bibitem[{West(2020)}]{west_calibration_2020}
West, R. 2020.
\newblock Calibration of Google Trends Time Series.
\newblock \emph{Proceedings of the 29th ACM International Conference on
  Information \& Knowledge Management} .

\end{thebibliography}
 }

\end{document}